\documentclass[aps,prl,reprint,superscriptaddress]{revtex4-1}
\usepackage{epic,eepic}
\usepackage{graphicx}% Include figure files
\usepackage{dcolumn}% Align table columns on decimal point
\usepackage{bm}% bold math
\usepackage{epstopdf}
\usepackage{amsmath}
\usepackage{amssymb}
\usepackage{latexsym}
\usepackage{revsymb}

\begin{document}

% v7 - ALG Comments of 2/1/17 incorporated
% v8 - shortened to make PRL word guideline (or at least, get close!)

%\preprint{}

\title{What is the temperature of an ultra-cold Rydberg plasma?}


%\author{}
%\thanks{}
%\affiliation{}
\author{Gabriel T. Forest}
\affiliation{Department of Physics and Astronomy, Colby College, Waterville, ME 04901, USA}
\author{Yin Li}
\affiliation{Department of Physics and Astronomy, Colby College, Waterville, ME 04901, USA}

\author{Edwin D. Ward}
\affiliation{Department of Physics and Astronomy, Colby College, Waterville, ME 04901, USA}

\author{Anne L. Goodsell}
\affiliation{Department of Physics, Middlebury College, Middlebury, VT 05753, USA}

\author{Duncan A. Tate}
\email[]{duncan.tate@colby.edu}
%\thanks{}
\affiliation{Department of Physics and Astronomy, Colby College, Waterville, ME 04901, USA}


\date{\today}

\begin{abstract}
We have measured the asymptotic expansion velocities of ultra-cold plasmas (UNPs) which evolve from cold, dense, samples of Rydberg rubidium atoms using ion time-of-flight spectroscopy. From this, we have obtained values for the initial plasma electron temperature as functions of the original Rydberg atom density and binding energy. Our results show that the electron temperature is determined principally by the plasma environment when the UNP decouples from the Rydberg atoms, which occurs when the plasma electrons become too cold to ionize the remaining Rydberg population. Furthermore, the dependence of the electron temperature on Rydberg atom density gives strong indirect evidence for the existence of a bottleneck in the spectrum of Rydberg states that coexist with a cold plasma.

\end{abstract}

%\keywords{}

\maketitle

%\section{}
%\subsection{}
%\subsubsection{}
The behavior and properties of ultra-cold neutral plasmas (UNPs) made by direct photoionization of cold atoms in a magneto-optical trap (MOT) are now relatively well understood (see for instance Ref. \cite{kill07}). Above the ionization threshold, $E_I$, conservation of linear momentum in the ionization process dictates that most of the excess photon energy, $\Delta E = h \nu - E_I$, goes to the electron. When the ionizing laser is tuned well above threshold, the initial electron temperature, $T_{e,0}$, is given by $\Delta E = \frac{3}{2} k_B T_{e,0}$. The asymptotic plasma expansion velocity is given by 
\begin{equation}
v_0 = \sqrt{k_B (T_{e,0} + T_{i,0})/m_i}, \label{vZero}
\end{equation} 
where $m_i$ is the ion mass, and $T_{i,0}$ is the initial ion temperature, determined largely by the temperature of the parent atoms to be of order $100 \ \mu$K. However, a number of mechanisms rapidly heat both electrons and ions. Specifically, close to threshold, three body recombination (TBR) heats the electrons and limits their temperature to a minimum value in the range 30 - 50 K. Additionally, at high density, continuum lowering (CL) will affect electron temperature \cite{hahn02}, and the ions are subject to disorder induced heating (DIH), which heats the ions to $\sim 1$ K in the first few microseconds of the plasma evolution process. As the plasma expands adiabatically on a time scale of order 10 - 100 $\mu$s, both the electron and ion temperatures fall, and the Coulomb coupling parameter, $\Gamma_\alpha$, increases \cite{kill07}, where
\begin{equation}
\Gamma_\alpha = \frac {e^2}{4 \pi \epsilon_0 a_{\alpha} k_B T_\alpha}, \label{gamma}
\end{equation}
and $a_{\alpha}$ is the Wigner-Seitz radius for species $\alpha$ (which may be electrons, $e$, or ions, $i$). %However, TBR, CL, and DIH limit $\Gamma$ values to $\Gamma_e \lesssim 1$ and $\Gamma_i \sim 10$ \cite{kill07}.
%, though there are proposals for increasing $\Gamma_e$ by adding cold Rydberg atoms to UNPs \cite{poh06a}, and for increasing $\Gamma_i$ by laser cooling ions (in UNPs made using alkaline earth atoms), multiple ionization, or via the dipole blockade \cite{muri15,bann13}.

On the other hand, the properties of plasmas which evolve from cold samples of Rydberg atoms (herein termed Rydberg plasmas) have not yet been studied as extensively as those formed by direct photionization, and in particular, the mechanisms which determine the electron and ion temperatures are not fully understood \cite{rob00,li04,rdsv13}. (We omit from this discussion the extensive work done on UNPs which evolve from cold Rydberg molecules in a supersonic beam, which have complex behaviors due to additional dynamical pathways \cite{schu16a}.) While it has been shown that dipole interactions between cold (``frozen'') Rydberg atoms play a significant role in the initial ionization \cite{tann08,robx14}, black body radiation \cite{bet09c,spen82a}, and collisions with hot background Rydberg atoms also contribute \cite{li04}. Once a critical electron density is achieved, an avalanche of electron-Rydberg collisions is initiated, and the plasma evolves mediated by the exchange of energy between the Rydberg atoms and the UNP. One consequence of dipole collisions contributing to the initiation of a UNP is that the potential energy so released is shared more equitably between electrons and ions than would be the case for direct photoionization. Additionally, the presence of a large reservoir of Rydberg atoms in the original target state, in nearby high-$\ell$ states, and in more deeply bound states during the initial stages of plasma evolution will favor different processes to those influencing photoionization-initiated UNPs, where the Rydberg states formed by TBR form a relatively weakly bound continuum which is in quasi-equilibrium with the plasma electrons. 

We have measured the effective initial electron temperature $T_{e,0}$ of UNPs which evolve from cold Rydberg samples by measuring their asymptotic expansion velocity $v_0$ from ion time-of-flight (TOF) spectra. The $nd_j$ Rydberg atoms ($24 \le n \le 110$) are excited from cold $^{85}$Rb atoms in a MOT which has a maximum atom density $1 \times 10^{10}$ cm$^{-3}$ ($1/\sqrt{e}$ radius $\sigma_0 \approx 400 \ \mu$m) and atom temperature $\sim 100 \ \mu$K. The atoms are excited to the $nd_j$ states using a narrow-bandwidth pulsed laser system (NBPL) which has been described elsewhere \cite{bran10}. The laser typically produces 5 ns duration, 50 - 100 $\mu$J pulses at 20 Hz, and has a line width of less than 200 MHz (we excite $j = {5/2}$ only for $n < 39$, but above this the $nd_{3/2}$ and $nd_{5/2}$ states are not resolved). Excitation of the cold atoms takes place between two high-transparency copper meshes separated by 18.3 mm which may be biased to null out external fields, and we can also apply voltage pulses to field ionize Rydberg atoms. We achieve Rydberg densities in the range $1 \times 10^7 - 1 \times 10^9$ cm$^{-3}$, and constantly monitor the number of atoms excited by measuring the 780 nm resonance fluorescence depletion when we field ionize the Rydberg atoms immediately after excitation \cite{han09}. (The densities have an absolute uncertainty of a factor of approximately 2, and a relative uncertainty of 20-30\%.) The NBPL laser beam is unfocussed, with a diameter of $\approx 4$ mm. This is much larger than the size of our cold atom sample, whose size we measure by imaging the 780 nm fluorescence onto a linear diode array.

The principle of the ion TOF technique is described in Ref. \cite{twed12}. After excitation, the cold Rydberg samples evolve to plasma over a period of $\le 10 \ \mu$s, which is negligible in comparison with the overall expansion time of the UNP (100 - 200 $\mu$s). The plasma slowly expands, and we detect plasma ions which exit the field-free interaction region between the meshes using a microchannel plate detector (MCP). The ions generally start to arrive at the MCP 30 - 40 $\mu$s after the NBPL pulse, and the TOF signal peaks between 80 and 100 $\mu$s, with an overall duration of $\le 400 \ \mu$s. We fit the ion TOF spectrum assuming a Gaussian density distribution with characteristic radius $\sigma = \sqrt{\sigma^2_0 + v^2_0 t^2}$ to extract a value for $v_0$. From $v_0$ we obtain $T_{e,0}$ using Eq. \eqref{vZero}, making the assumption that $T_{i,0} \lesssim 1$ K, and is therefore negligible in comparison to $T_{e,0}$. We have carried out extensive calibration of this technique by using it to find $v_0$ values for UNPs made by photoionizing cold atoms in the limit where $T_{e,0}$ is well above the regime in which TBR is important ($T_{e,0} = 50 - 300$ K), and find $v_0$ to be in agreement with Eq. \eqref{vZero}, if we ignore $T_{i,0}$ and use $\Delta E = \frac{3}{2} k_B T_{e,0}$. The values of $T_{e,0}$ we obtain for Rydberg UNPs lie in the range 20 - 120 K, with an estimated uncertainty $\sigma_{T_{e,0}} = \sqrt{(10 \ \textrm{K})^2 + (0.1 \times T_{e,0})^2}$. This uncertainty is primarily limited by the UNP density profile falling off more sharply than the Gaussian function assumed near the edges, and variations in the effective acceptance angle of the MCP. 

\begin{figure}
\centerline{\resizebox{0.50\textwidth}{!}{\includegraphics{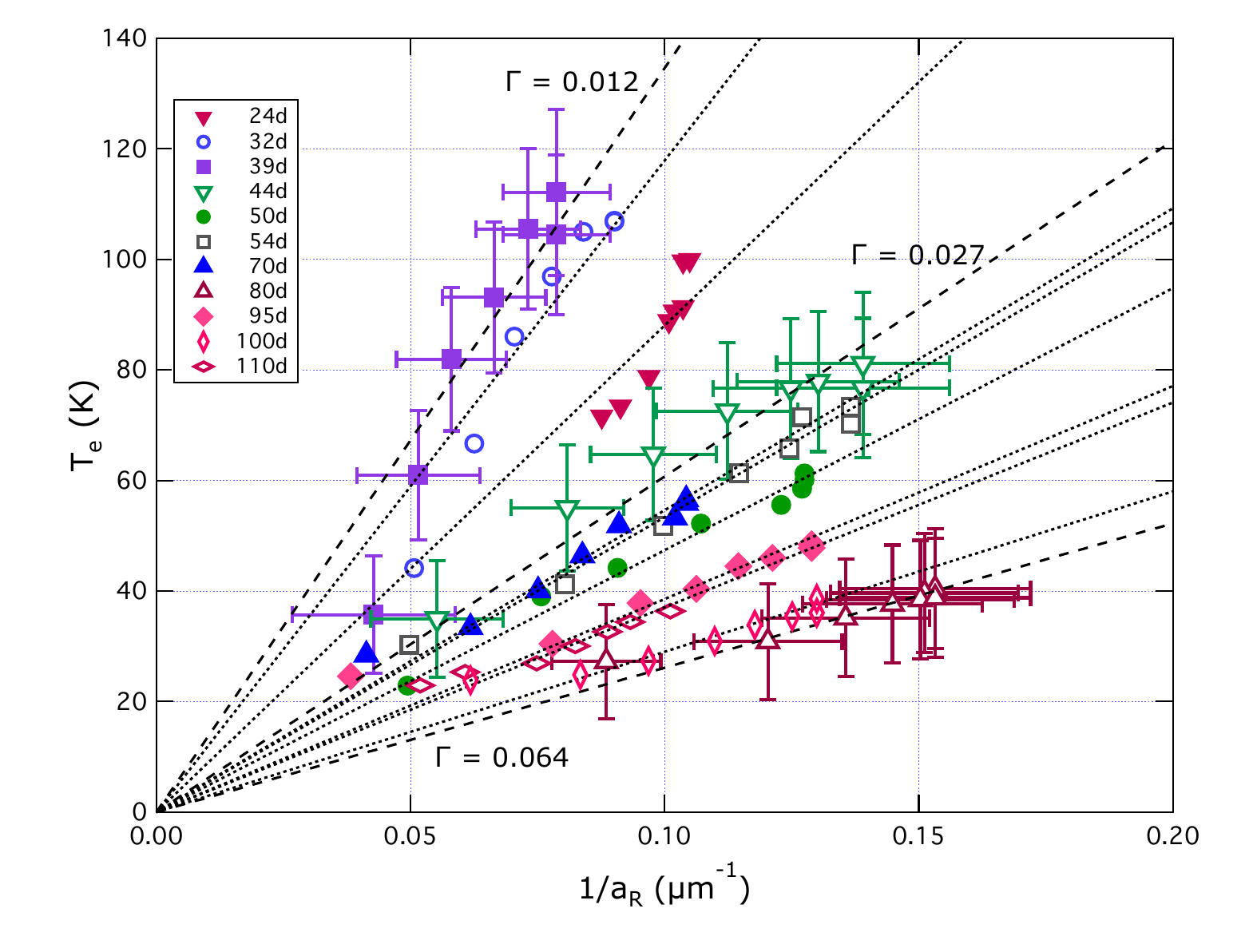}}}%{Fig1.eps}
\caption{Graph of $T_{e,0}$ versus $1/a_R$ for UNPs evolving from $nd$ cold Rb Rydberg samples in the range $24 \le n \le 110$. All the data for a given Rydberg state fall on a straight line whose $y$-intercept is zero within the uncertainties. The dashed/dotted lines are straight line fits that are constrained to have zero intercepts, and the corresponding $\Gamma_e$ values are given for $39d$, $44d$, and $80d$, assuming that $a_e = a_R$ (typical error bars are also given for these $nd$ states also). See text for details}
\label{TeVsa}
\end{figure}

Typical results for $T_{e,0}$ are shown in Fig. \ref{TeVsa} as a function of the reciprocal of the Rydberg atom spacing, $1/a_R = (4 \pi \rho_{avg}/3)^{1/3}$, where $\rho_{avg}$ is the average Rydberg atom density, for different $nd$ states in the range $24 < n < 110$. The low $T_e$, low $1/a_R$ ($n < 24$) cutoff in the data for a given Rydberg state is determined by the lowest density sample that would spontaneously evolve into a UNP, whereas the high $T_e$, high $1/a_R$ cutoff is determined by our maximum achievable density due to the declining oscillator strength of the $5p_{3/2} \rightarrow nd$ transition ($n > 110$). 

There are a number of interesting features in the data shown in Fig. \ref{TeVsa}. First, the results for a single Rydberg state fall on a straight line whose $y$-intercept is zero within the experimental uncertainty. The fraction of Rydberg atoms which ionize, $f$, will increase as $a_R$ decreases, so that the Wigner-Seitz radius for the electrons, $a_e$, is not a fixed multiple of $a_R$ ($a_e/a_R = 1/f^{1/3}$). This explains why data points for low-$n$ states at low $1/a_R$ (e.g. for $39d$) fall below the linear relationship - if plotted versus $1/a_e$, these points would be skewed to the left. The data shown in Fig. \ref{TeVsa} provide strong circumstantial evidence that the plasmas which form from a particular $nd$ state have approximately constant $\Gamma_e$. The corresponding $\Gamma_e$ values are shown on the graph, assuming that $f = 1 \Rightarrow a_e = a_R$. Using this assumption, we find that the values of $\Gamma_e$ vary from 0.01 ($39d$) to greater than 0.06 ($80d$). These values are comparable with those reported in experiments on UNPs made by direct photoionization \cite{flet07,gupt07}. Additionally, we have compared our $\Gamma_e$ values with those obtained from a Monte-Carlo model provided to us by Francis Robicheaux \cite{robx02,robx03}. While the model analyzes UNPs made by photionization, rather than those which evolve from Rydberg samples, one expects there should be a reasonably smooth variation in the plasma properties in the region of the ionization limit. When we run this code using density and size parameters comparable to our experiment $\rho_{avg} \sim 10^8$ cm$^{-3}$ and $\sigma_0 \approx 400 \ \mu$m, we find $\Gamma_e$ values in the range 0.01 ($T_{e,0} = 140$ K) to 0.09 ($T_{e,0} = 20$ K). 

\begin{figure}
\centerline{\resizebox{0.50\textwidth}{!}{\includegraphics{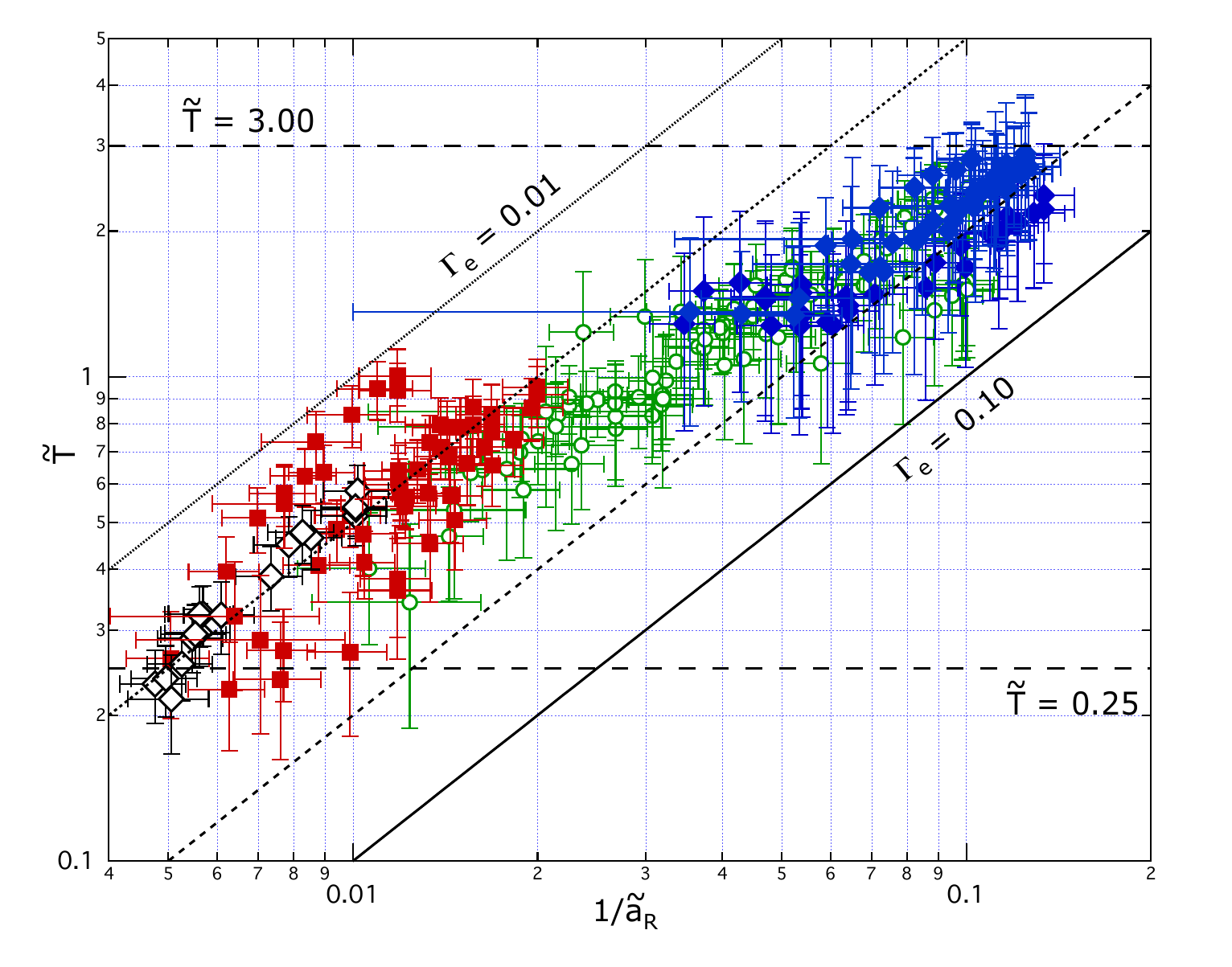}}}%{Fig1.eps}
\caption{Graph of $\tilde T$ versus $1/\tilde a_R$ for UNPs evolving from $nd$ cold Rb Rydberg samples in the range $24 < n < 120$. The data shown comprise 24 different $n$ values, and for each $n$, data were obtained for between six and 20 different densities. We distinguish the data in terms of different binding energies: $E_b > 200$ K (open black $\lozenge$); $200 > E_b > 100$ K (filled red $\blacksquare$); $100 > E_b > 25$ K (open green \begin{large}$\circ$\end{large}); and $E_b < 25$ K (filled blue $\blacklozenge$). Also shown are lines corresponding to $\tilde T = 0.25$ and $\tilde T = 3.0$, and (from top to bottom) the lines corresponding to $\Gamma_e = 0.01, \ 0.02, \ 0.05$, and $0.1$, found assuming that $a_e = a_R$.}% See text for details.}
\label{scaledTvsscaleda}
\end{figure}

The second feature apparent in Fig. \ref{TeVsa} is that $\Gamma_e$ generally increases as the magnitude of the binding energy, $E_b$, decreases, where $E_b = \frac{1}{2} \frac{e^2}{4 \pi \epsilon_0 a_0} \frac {1}{{n^\ast}^2}$ (in which $a_0$ is the Bohr radius and $n^\ast$ is the effective principal quantum number of the initial $nd$ state, $n^\ast \approx n - 1.35$). The apparent correlation between $\Gamma_e$ and the initial Rydberg state binding energy, $E_b$, suggests that it would be useful to plot the data using the scaled quantities $\tilde T = k_B T_{e,0}/E_b$ and $\tilde a_e = a_e/2{n^\ast}^2a_0$. Using this scaling, it can be seen that Eq. \eqref{gamma} for electrons can be expressed as
\begin{equation}
\Gamma_e = \frac {1}{\tilde a_e \, \tilde T}, \label{gammaScaled}
\end{equation}
and thus for constant $\Gamma_e$, one expects $\tilde T \propto 1/\tilde a_e$. We have therefore plotted $\tilde T$ versus $1/\tilde a_R$, as shown in Fig. \ref{scaledTvsscaleda}. As can be seen, scaling the data in this way results in a single universal curve. 

As in Fig. \ref{TeVsa}, the ratio of the Wigner-Seitz radius for the electrons in the UNP to the mean Rydberg atom separation varies with $f$. However, theoretical results given in Refs. \cite{robx02, robx03,poh03} show that a maximum of $f \approx 0.7$ of the Rydberg atoms ionize during the avalanche for principal quantum numbers in the range $n = 45 - 70$ and densities of $10^8 - 10^9$ cm$^{-3}$. This suggests that, for the data with $1/\tilde a_R > 0.1$, $\tilde T > 2$ in Fig. \ref{scaledTvsscaleda}, $1/\tilde a_e = f^{1/3}/\tilde a_R \approx 0.9/\tilde a_R$. For for these data points, the values of $\Gamma_e \approx 0.06$ found assuming $a_e = a_R$ are therefore quite accurate. On the other hand, in the low $1/\tilde a_R$, low $\tilde T$ part of the graph, the data points would be skewed to the left relative to those in Fig. \ref{scaledTvsscaleda} when plotted versus $1/\tilde a_e$, and the lines of constant $\Gamma_e$ will behave differently than shown in Fig. \ref{scaledTvsscaleda}. 
%(However, due to the $f^{-1/3}$ scaling, these differences have relatively weak dependence on the ionization fraction: $a_e/a_R \approx 2$ for $f = 0.1$ and $a_e/a_R \approx 5$ for $f = 0.01$.)
% Got to here 1/27/17
%[Low $1/\tilde a_R$: if $f=0.1$, $1/\tilde a_e \approx 0.46/\tilde a_R$]

While our experiment does not specifically address the initial ionization mechanism which seeds the plasma, other experiments \cite{tann08, rdsv13} have observed fast, $\le 1 \ \mu$s, plasma formation from Rb Rydberg samples with densities $\lesssim 10^{11}$ cm$^{-3}$ for $n \approx 50$, and these works identified dipole interactions as the initial seeding mechanism. However, theoretical modeling of cold dipole collisions does not reproduce the rate of ionization seen in the experiments \cite{robx14}. Additionally, in our experiment, one can estimate the ionization fraction needed to trap electrons with temperature $E_b/k_B$ using the depth of a potential well of $N_i$ positive ions in a Gaussian density distribution with radius $\sigma_0$, i.e, $U = \sqrt{2/\pi} \, N_i \, e^2/4 \, \pi \, \epsilon_0 \, \sigma_0$. Setting $U = E_b$ gives $N_i \approx 9,000$ for $24d$, or an ionization fraction of $6 \times 10^{-4}$ for the atom density we achieved for this state, and $N_i \approx 500$ for $100d$, or an ionization fraction of $4 \times 10^{-4}$. Calculations based on the theoretical results reported in \cite{robx14} predict ionization probabilities for dipole-mediated collisions of cold Rydberg atoms of $6 \times 10^{-5}$ at $n=24$ and $3 \times 10^{-3}$ at $n=100$ for the respective $\tilde a_R$ values we can attain. So, while cold dipole collisions can thus clearly provide enough ionization to establish a plasma for $n=100$, the predicted ionization fraction at $n=24$ is 10 times too low to reach the threshold condition. Some other mechanism must dominate purely cold collisions at low $n$, either collisions of thermal Rydberg atoms from the background vapor \cite{li04}, or ionization caused by blackbody radiation (BBR) \cite{poh03}. Furthermore, according to the theoretical results discussed above, the dominant seeding mechanism changes somewhere between $n=40$ and $70$. This range corresponds to the open green circles in Fig. \ref{scaledTvsscaleda}, and since we see no obvious discontinuity here we conclude that the effective electron temperature of our plasmas is not strongly dependent on the seeding mechanism. 

Rather, the data in Fig. \ref{TeVsa} show that the electron temperature of a Rydberg plasma is determined by the equilibrium established between the free electrons in the UNP and the reservoir of Rydberg states that they scatter from during the avalanche regime. Specifically, $T_{e,0}$ increases with the Rydberg density (increasing $1/a_R$) for a given Rydberg state, and in general, at a given density, $T_{e,0}$ is larger for atoms whose initial binding energy is larger. This is characteristic of electron-Rydberg collisions in which the Rydberg atom is deexcited, and the electron temperature increases: the energy given to the electrons is proportional to the Rydberg binding energy \cite{robx03}. Thus, while collisions between low-$n$ state Rydberg atoms and electrons have a lower cross section than for high-$n$ states, and are therefore less probable at comparable densities, the plasma heating from electron collisions with low-$n$ state atoms is much more significant. This behavior is manifested in the data in Fig. \ref{TeVsa}, but is more obvious in Fig. \ref{scaledTvsscaleda} - Rydberg states with larger binding energy (lower $n$) are hotter than those with smaller binding energies (higher $n$), but $\tilde T$ is a smaller for low-$n$ states than for high-$n$ states. 

The results shown in Figs. \ref{TeVsa} and \ref{scaledTvsscaleda} thus imply that the most significant factor that determines the plasma electron temperature as a fraction of the binding energy is the number of electron collisions with the reservoir of Rydberg atoms, and must therefore be related to $f$, the fraction of the Rydberg atoms which ionize during the avalanche. One can estimate the effect of $f$ on the plasma temperature by considering the correlation between the ``bottleneck'' Rydberg binding energy and the temperature of the plasma electrons with which the Rydberg atoms are in equilibrium. It has been shown in numerous theoretical studies that there exists a bottleneck energy, $E_{bn}$, of the Rydberg state distribution which is populated by recombination of electrons and ions in a plasma with electron temperature $T_e$ \cite{mans69,kuz02b,robx03,poh08,bann11}. Specifically, $E_{bn} \approx 4k_BT_e$, and Rydberg states with binding energy $E_b < E_{bn}$ will eventually ionize, while those with $E_b > E_{bn}$ will be deexcited to states with lower $n$ which will eventually decay radiatively to the ground state. 

%Unfortunately, it isn't experimentally possible to observe the Rydberg states at the bottleneck energy directly. The only feasible method to do so is selective field ionization (SFI) \cite{gall94}. However, since electron-Rydberg collisions populate a large range of angular momentum states, $\ell$ for each $n$, and each $n,\ell$ state has many adiabatic and non-adiabatic field ionization thresholds, any peak in the Rydberg energy spectrum observed using SFI would be smeared out over a very wide range of field strengths. Consequently, available experimental evidence for the bottleneck is circumstantial: the calculated rate constants for excitation, deexcitation, and ionization in electron-Rydberg collisions on which the existence of the bottleneck depends have been used to model the evolution of UNPs made by photoionization, and the predictions compared with experiment, e.g., in Refs. \cite{gupt07,poh08}.

We use the mechanism for the evolution of cold Rydberg atoms to plasma reported in the theoretical analyses of Refs. \cite{robx03} and \cite{poh03} to make an estimate of the effective initial electron temperature, $T_{e,0}$ of a UNP formed by an initial population of cold Rydberg atoms with binding energy $E_{b,i}$. Our analysis assumes the existence of the bottleneck, and accurately predicts the range of $\tilde T$ we see in our results. As described in Refs. \cite{robx03} and \cite{poh03}, the plasma forms by ionization of Rydberg atoms by dipole-mediated collisions of cold atoms, or by ionization by BBR, or by collisions between the small fraction (1-2\%) of thermal (300 K) Rydberg atoms and cold Rydberg atoms. The latter mechanism will produce fast electrons, which escape from the interaction region. However, both cold-cold collisions and BBR produce slower electrons with energy $k_BT_e \sim E_{b,i}$. In the theoretical work reported in Ref. \cite{robx05}, it was shown that cold dipole collisions ionize a small fraction of the Rydberg atoms, producing free electrons with average energy $E_{b,i}$, while the deexcited Rydberg atoms have a distribution of binding energies, but all are bound by at least $2E_{b,i}$. Additionally, the cross section for direct photoionization by BBR falls rapidly above the ionization limit, resulting in electrons with energy $\ll k_BT_{BBR}$, where $T_{BBR}$ is the temperature of the BBR, $\approx 300$ K \cite{spen82a,bet09c}. %Unfortunately, both papers concentrate on the Rydberg state distributions, and do not consider the electron temperature in the UNP. (Also, no systematic dependence on binding energy was explored.) 

When enough atoms have ionized that low-energy electrons are trapped, a plasma forms. At this instant, at the onset of the avalanche, the electron temperature is more than sufficient, $T_e > 0.25 \times E_{b,i}/k_B$, to ionize the Rydberg atoms in the original state, as well as many of the partner atoms deexcited by cold dipole collisions. The interplay between electron-Rydberg exciting, deexciting, and ionizing collisions, and recombination maintains the energy balance in the evolution so that the average binding energy of the un-ionized atoms increases (i.e., the population distribution migrates to lower $n$). Naively, this process should end when the final average binding energy, $E_{b,f}$, becomes greater than the bottleneck energy, $4k_BT_{e}$, since at this point the electrons in the plasma can no longer ionize the Rydberg atoms. On the other hand, the bottleneck condition is based on probability flux, not on energy transfer. When a Rydberg atom is deexcited in a collision with an electron, the energy given to the plasma is proportional to the binding energy, while the energy lost in a collision which excites the atom is proportional to $k_BT_e$ \cite{robx03}. Therefore, it isn't clear that $E_{b,f} = 4k_BT_{e}$ at the decoupling point. In fact, at this limit, it is likely that the heating collisions dominate cooling collisions, suggesting that $E_{b,f} < 4 k_BT_e$. All we can say is that the avalanche ends and the plasma and the Rydbergs decouple when $E_{b,f} = \beta k_BT_{e}$, where $\beta \approx 4$. Thereafter, the UNP cools as it expands adiabatically with an asymptotic expansion velocity, $v_0$, which will be characteristic of temperature $T_{e,0}=T_e=E_{b,f}/\beta k_B$. There will still be some residual recombination which populates high-$n$ states, but the amount of heat they add to the plasma is small relative to that converted to outward expansion, and they, too, progressively decouple from the plasma (i.e., the minimum $n$ that will ionize in a collision with an electron increases with time as the plasma expands) \cite{poh03}. %(since $E_{b,f} > k_BT_{e,0}$)

We can use this scenario to estimate the relation between $T_{e,0}$ and $E_{b,i}$ and show that the data in Fig. \ref{scaledTvsscaleda} provide strong evidence for the existence of the bottleneck. Ignoring energy added to the system by BBR and hot-cold Rydberg collisions, the initial energy of the system is $-N_R \, E_{b,i}$, and the final energy (before adiabatic expansion begins) is $-(1-f) \, N_R \, E_{b,f} + f \, N_R \, (3/2) \, k_B \, T_{e,0}$, where $N_R$ is the initial number of Rydberg atoms. Equating these energies, and using $E_{b,f} = \beta k_BT_{e,0}$, we obtain the following relationship
\begin{equation}
\tilde T = \frac{k_BT_{e,0}}{E_{b,i}} = \biggl [(1-f)\beta - \frac{3}{2}f \biggr ]^{-1}. \label{ScaledT}
%\tilde T = \frac{k_BT_{e,0}}{E_{b,i}} = \biggl [\beta-(\beta + \frac{3}{2}) \, f \biggr ]^{-1}. \label{ScaledT}
\end{equation}
This analysis is over-simplified for a number of reasons, the main one being that it ignores any dependence of $\beta$ on $f$. Nevertheless, it predicts boundary values for $\tilde T$ which match our data well. Using $\beta = 4$, one obtains $\tilde T = 0.25$ for $f = 0$, and $\tilde T = 3.0$ for $f=0.67$ (these limits are shown in Fig. \ref{scaledTvsscaleda}). In the former case, which corresponds to large $E_{b,i}$, the UNPs barely form - some laser shots populate enough Rydberg atoms that the small fraction which ionizes is sufficient to trap subsequent electrons, but some do not. However, in the plasmas which do form, the electrons are hot and the UNP expands quickly, and decouples from the Rydberg reservoir early. During the brief avalanche, the change in the Rydberg population is minimal, and the electrons equilibrate with a large reservoir of atoms with average energy which is only slightly less than $E_{b,i}$. Hence, the dominant interaction is one which cools the electrons until $\beta k_BT_{e,0} \approx E_{b,i}$. As can be seen in Fig. \ref{scaledTvsscaleda}, the low-$n$ states (which correspond to small $1/\tilde a_R$) have $\tilde T \approx 0.25$, an experimental result which clearly depends on the existence of the bottleneck and is not materially affected by the simplicity of the model. On the other hand, states with low $E_{b,i}$ (which ionize easily at large $1/\tilde a_R$) have an initial electron energy which is small, $\approx E_{b,i}$. Additionally, since the electron-Rydberg collision cross section is large, a larger fraction of the Rydberg atoms are ionized, and a larger multiple of $E_{b,i}$ is ``mined'' by the electrons to heat the plasma. Unfortunately, the upper limit of $\tilde T$ predicted by our model is not very precise. Specifically, there is a singularity predicted by Eq. \eqref{ScaledT} at $f = \beta/(\beta + 3/2)$ ($f = 0.73$ for $\beta = 4$). However, in this regime is reached close to the ionization limit ($E_{b,i}/k_B \lesssim 10$ K) where continuum lowering \cite{hahn02} invalidates the model's assumptions anyway. On the other hand, in the theoretical analysis shown in Fig. 3 in Ref. \cite{poh03}, the authors found $E_{b,f}/E_{b,i} = \beta k_B T_{e,0}/E_{b,i} = \beta \tilde T \approx 8$ when $f \approx 0.7$ and $E_{b,i}/k_B \approx 30$ K ($n_i = 70$). While the analysis was done for a significantly higher density than we can achieve, these values clearly imply that our observed upper limit value of $\tilde T = 3$, obtained assuming $\beta = 4$ and $f = 0.67$, is reasonable (for the $70d$ state, we observed a maximum $\tilde T$ of $1.7\pm0.3$, which gives $\beta \tilde T = 6 \pm 1$ using $\beta = 4$). %On the other hand, the value of $E_{b,f}/E_{b,i} \approx 8$ suggests that a value of $\beta$ which is less than 4 is more appropriate, at least at higher densities. 

In summary, our experiment shows that the electron temperature in a cold plasma which evolves from a dense sample of cold Rydberg atoms is determined principally by the last point of interaction between the UNP and the remaining deexcited Rydberg atoms. Furthermore, the dependence of the electron temperature on Rydberg atom density gives strong indirect evidence for the existence a bottleneck in the spectrum of Rydberg states in coexistence with a cold plasma.

We acknowledge extensive discussions with Francis Robicheaux, Tom Gallagher, and Charlie Conover. This work has been supported by Colby College, via the Division of Natural Sciences grants program, by Middlebury College, and by NSF (grant number PHY-1068191).

\raggedright
\bibliography{PRL_Bib_092116}

\end{document}